\documentclass{aa}  

\usepackage{graphicx}
\usepackage{txfonts}
\usepackage[breaklinks,colorlinks,linkcolor=blue,citecolor=blue,urlcolor=blue]{hyperref}
\usepackage{natbib}

\usepackage{siunitx}
\DeclareSIUnit{\solarmass}{M_\sun}
\DeclareSIUnit{\solarlum}{L_\sun}
\DeclareSIUnit{\year}{yr}
\DeclareSIUnit{\parsec}{pc}
\DeclareSIUnit{\fluxunit}{erg~cm^{-2}~s^{-1}}
\DeclareSIUnit{\fluxdensityunit}{\fluxunit~\angstrom^{-1}}
\DeclareSIUnit{\fluxdensityunitperpixel}{\fluxunit~\angstrom^{-1}~subpix^{-2}}
\DeclareSIUnit{\fluxdensityunitperarcsec}{\fluxunit~\angstrom^{-1}~arcsec^{-2}}
\makeatletter
\renewcommand*\aa@pageof{, page \thepage{} of \pageref*{LastPage}}
\makeatother

\begin{document}

   \title{Planetary Nebulae with UVIT II: Revelations from FUV vision of Butterfly Nebula NGC~6302\thanks{Based on data obtained with Ultra-Violet Imaging Telescope (UVIT) on satellite ASTROSAT }}

      \author{N. Kameswara Rao \inst{1}\and O. De Marco \inst{2}\and S. Krishna \inst{1}\and J. Murthy\inst{1}\and A. Ray\inst{3}\and F. Sutaria \inst{1}\and R. Mohan \inst{1}}

   \institute{Indian Institute of Astrophysics, Koramangala II Block, Bangalore-560034, India\\
   \email{nkrao@iiap.res.in}
   \and
   Macquarie University, NSW, 2109, Australia
   \and
   Homi Bhabha Centre for Science Education (TIFR), Mumbai-400088, India
   }

   \date{Received May 26, 2018; accepted October 3, 2018}
 
  \abstract{
    The high excitation planetary nebula,  NGC~6302,  has been imaged in two
 far-ultraviolet (FUV) filters, F169M (Sapphire; $\lambda_{\rm eff}$: {1608~\AA})  and F172M (Silica; $\lambda_{\rm eff}$: {1717~\AA})  and two NUV filters, N219M (B15; $\lambda_{\rm eff}$: {2196~\AA}) and N279N (N2; $\lambda_{\rm eff}$: {2792~\AA})  with the Ultra Violet Imaging 
 Telescope (UVIT). 
 The FUV F169M image shows faint emission lobes that extend to about
 5~arcmin on either side of the central source. Faint orthogonal collimated
 jet-like structures  are present
 on either side of the FUV lobes through the 
 central source. These structures are not present in the two 
 NUV filters nor in the FUV F172M filter. 
  Optical and IR images of NGC 6302 show
 bright emission bipolar lobes in the east-west direction with a massive torus
 of molecular gas and dust seen as a dark lane in the north-south direction.
 The FUV lobes are much more extended and oriented at a position angle
 of 113$^\circ$.
 They and the jet-like structures might be remnants of an earlier
  evolutionary phase, prior to the dramatic explosive event that
 triggered the Hubble type bipolar flows approximately 2200~years ago.
 The source of the FUV lobe and jet emission is not known, but is likely due to
 fluorescent emission from H$_{2}$ molecules. The cause of the  difference in orientation of
 optical and FUV lobes is not clear and, we speculate, could be related to two binary interactions. }

   \keywords{Stars: AGB and post-AGB -- Stars: winds, outflows -- planetary nebulae: general -- planetary nebulae: individual: NGC~6302}
   \titlerunning{FUV vision of Butterfly Nebula NGC~6302}
   \authorrunning{N.K.Rao et~al.}
   \maketitle

\section{Introduction}
\label{sec:intro}
\object{NGC~6302} (\object{PN~G349.5+01.0}) is a typical bipolar (multipolar?)
planetary nebula (PN) discovered by \citet{barnard1906} as early as 1880 using the 
36-inch Lick refractor. He called it the `bug nebula'. It is now
well known as the Butterfly Nebula.
Very detailed Hubble Space Telescope (HST) images have been
discussed by \citet{szyszka09}, who also identified the elusive central star.
The optical narrow-band images show two main lobes with a complicated clumpy 
small-scale structure in the east-west direction, separated by a dark lane of 
very dense gas (neutral and  molecular) and
dust, stretching in the north-south direction.  The gas and dust formed into  a toroid,
 which obscures the
 central star with a visual extinction of about 8 magnitudes \citep{matsuura05,peretto07,szyszka09,wright11}.
 \citet{meaburn08}
 determined the distance to the nebula to be \SI{1.17(14)}{\kilo\parsec} from its expansion 
 parallax found using proper motions of features in the north-west lobe. This 
 estimate seems to be consistent with measurements of proper motions from
 HST images of the eastern lobe \citep{szyszka11}.      
 \citet{wright11} used 3D photoionization modelling of the nebula to
 derive  the properties of the central star. They found it to be hydrogen deficient 
 with  T$_{\rm eff}$ of \SI{220000}{K}, log $g$ of 7, L$_{\rm *}$ of 
 \SI{14300}{\solarlum} and a mass of \SIrange{0.73}{0.82}{\solarmass}, with an initial
 mass estimated to be around \SI{5.5}{\solarmass}.

 The nebula has been classified as a Type I PN \citep{peimbert83,kingsburgh94}, implying high helium and nitrogen
 abundances. \citet{wright11} redetermined the abundances using emission 
 lines from both the lobes that range in ionization from [\ion{O}{i}]  to 
 [\ion{Si}{ix}] and concluded that the high helium (more than that expected for a Type I
 PN) and very high nitrogen abundance (that showed the sum of C,N,O to be  much 
 larger than the solar value) are a result of the precursor star undergoing 
 a third dredge-up in its earlier evolution.

 Extensive studies of the circumstellar torus from infrared to 
 radio wavelengths \citep{lester84,kemper02,matsuura05,peretto07,santandergarcia17}
 suggest that the structure is a broken disc containing \SI{2.2}{\solarmass} of
 dust and molecular gas expanding with a velocity of \SI{8}{km.s^{-1}} that 
 was ejected by the star some \num{5000} years ago, over a period of about \num{2000} years.
 The torus also obscures both the star and a smaller ionized gas disc (seen 
 in \SI{6}{cm} free-free continuum) around the star. Kinematic studies of the east-west
 lobes seem to suggest that an explosive event initiated a kind of Hubble 
 flow (velocity increases outward proportional to the distance away from the
 star) in both lobes about \num{2200}~years ago \citep{meaburn08,szyszka09}.
 
The kinematic history of this nebula is complex: according to the modelling by \citet{santandergarcia17} the bulk of the material was ejected and shaped into a ring of material and a set of lobe fragments in an event starting \SI{5000}{\year} ago. Optical lobes were ejected after a brief delay between \SIrange{3600}{4700}{\year} ago, followed by the large and fast-expanding north-west lobe that was ejected around \SI{2200}{\year} ago. The new inner ring (in the torus) was ejected approximately same time as the NW lobe, \SI{2200}{\year} ago. As is often the case in collimated nebulae like this one, it cannot be clearly stated whether the bipolar outflow was a result of disk collimation or whether it resulted from jet-like launching. In this case, the disk may have been able to collimate the lobes as opposed to, for example, the case of \object{Mz~3} \citep{smith05,macdonald18}.

 Several possibilities have been suggested for the formation of 
 bipolar nebulae involving single and double stars. In case of single stars
 equatorial discs in the advanced AGB phase
 could shape them after the emergence of fast stellar wind \citep{balick02,icke03}.
 However, this suggestion has  an
 inherent problem of lack of enough angular momentum \citep{soker06,nordhaus06,garciasegura14}. In case of 
 binary stars, the companion can collimate the AGB ejecta into a
 bipolar
 structure \citep{soker98}. However it is not clear whether all bipolar 
 nebulae are shaped by binary star interactions.
   In the case of \object{NGC~6302}, \citet{peretto07} and \citet{wright11} provided arguments that the mass loss rate of 
\SI{5e-4}{\solarmass\per\year} required to produce the massive torus 
 cannot be 
 driven by a single star, but it could easily be accomplished by a binary star.
 However, no  binary companion has been detected so far 
 in the \object{NGC~6302} system. The requirements to produce a NGC~6302-kind system
 seem to demand  a hydrogen deficient \SI{220000}{K} hot post-AGB  star of
 \SIrange{0.7}{0.8}{\solarmass} with a companion (presently unseen). 
 The primary must have gone through a third dredgeup phase, and
 initially  ejected  \SI{2}{\solarmass} of gas and dust in the equatorial 
(or orbital) plane about \SI{5000}{yr} ago. It later, produced 
 a Hubble-type flow through polar lobes by an explosive event (the nature of 
 which is unknown) \SI{2200}{yr} ago. 
 How such a system is produced in the course of stellar 
 evolution of a single or double star is not clear.

         In the present paper we report the discovery of faint FUV lobes, much
 more extended than the optical lobes, with two faint jets in the
 orthogonal directions on either side of the lobes, as seen in some 
interacting binary systems. Surprisingly the FUV lobes are not oriented
 parallel to the optical lobes (and torus). This may
 provide clues to the past evolution of the system. The paper is structured
 as follows:  this introduction is followed by our UVIT observations (Sect.~\ref{sec:obs}). The results  are presented in Sect.~\ref{sec:res} and  discussed in Sect.~\ref{sec:disc}. Finally we offer some
 concluding remarks in Sect.~\ref{sec:concl}.

\section{Observations}
\label{sec:obs}
\begin{figure*}
\centering
\includegraphics[width=18cm,height=18cm]{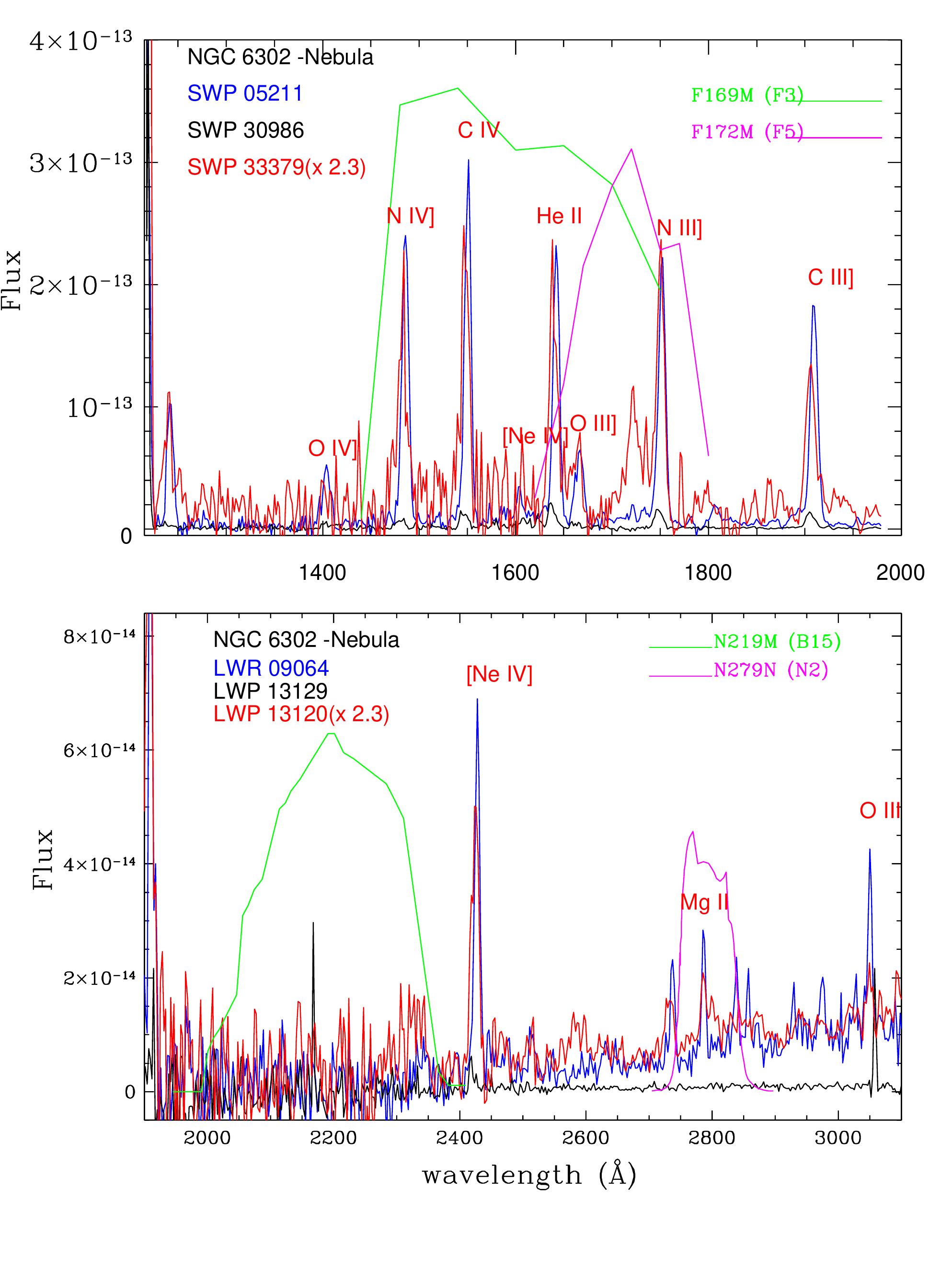}
\caption{ IUE low resolution nebular spectrum of \object{NGC~6302} (blue line). The FUV range is 
  plotted in the top panel and the
  NUV range in the bottom panel with UVIT filter effective areas (relative) shown.
  Top panel: the green
 curve is filter F169M (Sapphire;
  $\lambda_{\rm eff}$: \SI{1608}{\angstrom})  and the filter F172M (Silica;
$\lambda_{\rm eff}$: \SI{1717}{\angstrom})  is shown in magenta.
   Bottom panel: 
the green curve denotes the NUV filter N219M (B15;
  $\lambda_{\rm eff}$: \SI{2196}{\angstrom}) while the filter N279N (N2;
 $\lambda_{\rm eff}$: \SI{2792}{\angstrom}) is shown in  magenta.
  The IUE spectra  obtained on the central position \citep{feibelman01} is shown in blue and
  offset positions are shown in  red and black. Line identifications
 are from \citet{feibelman01} and \citet{beintema99}. The flux of the red spectrum has been
 multiplied by a constant factor of 2.3 to scale it with the nebular spectrum from the centre on the [\ion{N}{iv}] line at \SI{1486}{\angstrom}.}
\label{fig:iuespectrum}
\end{figure*}

Imaging observations of \object{NGC~6302} were obtained with UVIT on 2017 March
 18 in four UV filters. UVIT is one of the five payloads on the 
multi-wavelength
 Indian astronomical satellite ASTROSAT that was launched on 2015 September 28.
 It consists of two 38-cm aperture telescopes, one of which is optimized for
 the FUV; the other, with a dichroic beam splitter that reflects the NUV and transmits
 the optical. UVIT provides images in three channels: FUV, NUV and optical
 simultaneously, with a 28-arcmin diameter field of view. The best spatial resolution in the
 UV bands is about 1.3~arcsec. Each of the UV channels includes 
 five filters and one low resolution transmission grating. The visible channel is
 used for tracking, with limited scientific utility.  Details of the instrument are provided in
 \citet{kumar12} and the in-orbit performance is described in 
 \citet{tandon17a}.

         The present observations of \object{NGC~6302} were obtained in two FUV filters:
 F169M (Sapphire;
  $\lambda_{\rm eff}$: \SI{1608}{\angstrom})  and F172M (Silica;
$\lambda_{\rm eff}$: \SI{1717}{\angstrom})  as well as in two NUV filters: N219M (B15;
  $\lambda_{\rm eff}$: \SI{2196}{\angstrom}) and N279N (N2; $\lambda_{\rm
 eff}$: \SI{2792}{\angstrom}). The effective exposure times that went into
 constructing the images in various filters are the following: \SI{686}{s}  in F169M, \SI{622}{s} in F172M, \SI{930}{s} in N219M and \SI{509}{s} in the N279N. The stellar images in
 N279N and F169M filters show a PSF of 1.4 arcsec.
 These observations are complemented with weak exposures in the F154W (BaF${\rm 2}$:
$\lambda_{\rm eff}$ of \SI{1541}{\angstrom}) and NUV grating.  

        In its standard operating mode, UVIT will take images of the sky with a frame rate of \SI{29}{s^{-1}}, which are stored on board and then sent to the
 Indian Space Science Data Centre (ISSDC), where the data are written into 
instrument specific Level 1 data files. \citet{murthy17} have written a set
 of procedures (JUDE) to read the Level 1 data, extract the photon events
 from each frame, correct for spacecraft motion (image registration) and add
 into an image. We used astrometry.net \citep{lang10} for an astrometric
 calibration and \citet{rahna17} and \citet{tandon17b}, for a photometric
 calibration. We have
 co-added the individual images and placed them all on a common reference frame
  and these were used for our scientific analysis.

\begin{figure*}
\centering
\includegraphics[width=\textwidth]{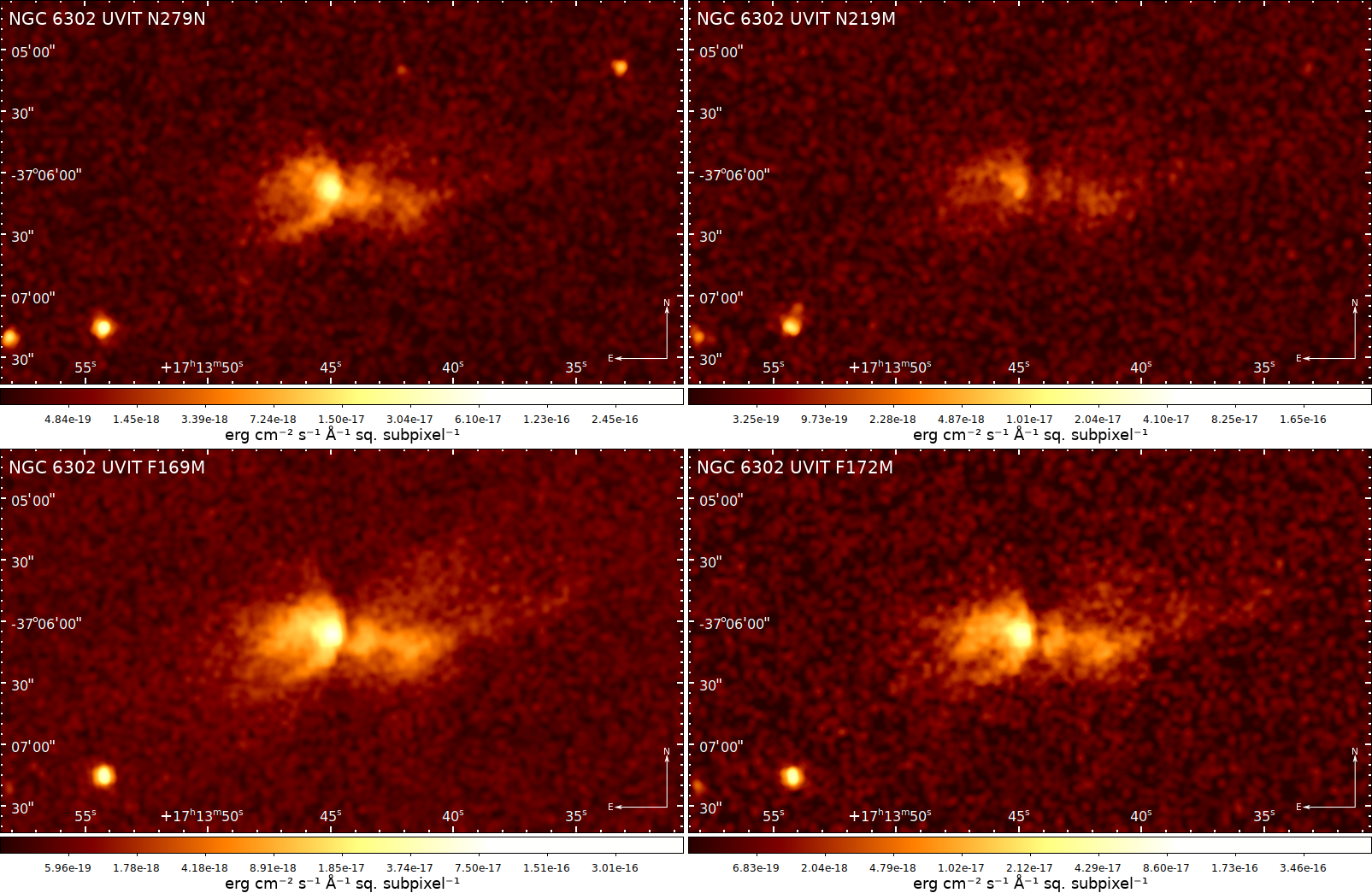}
\caption{UVIT images of \object{NGC~6302}. Top panel: NUV  image in the filter  
 N279N on the left and the N219M filter on the right. Bottom panel: FUV
 images of the filter F169M on the left
 and  images in the filter F172M on right.}
\label{fig:imagecomp}
\end{figure*}
Figure \ref{fig:iuespectrum} shows the IUE spectra of \object{NGC~6302} in the FUV (short wavelength; top panel) and the NUV (long wavelength; bottom panel) at three positions
 in the nebula: the centre of the nebula (blue line - SWP 05211 and LWR 09064), 
 offset by \SI{4}{\arcsec} east and \SI{9}{\arcsec} south (red line) and \SI{12}{\arcsec} east and \SI{6}{\arcsec} north (black line). 
 The emission
  lines decrease in strength appreciably away from the centre.
    The transmission functions of the
  filters (on an arbitrary scale) used are shown in green (F169M, N219M) and magenta  (F172M, N279N) superposed on the IUE spectra. The F169M filter has a passband of \SI{290}{\angstrom}
 and includes mainly high excitation lines due to [\ion{N}{iv}] at \SI{1486}{\angstrom},
  \ion{C}{iv} at \SI{1550}{\angstrom}, [\ion{Ne}{iv}] at \SI{1601}{\angstrom}, \ion{He}{ii} at \SI{1640}{\angstrom}, 
 [\ion{O}{iii}] at \SI{1663}{\angstrom} and \ion{N}{iii}] at \SI{1760}{\angstrom} at the centre of the nebula
  whereas the F172M filter, which has a passband of \SI{125}{\angstrom}, includes only 
 \ion{N}{iii}] at \SI{1760}{\angstrom}. However, the NUV filter N219M  has a passband of
 \SI{270}{\angstrom}, centred on the ISM absorption bump at \SI{2179}{\angstrom} and does not include
  any strong emission lines. The IUE spectrum of the nebula, as noted by \citet{feibelman01}, does not show any
 interstellar absorption bump although a reddening of c(H$\beta$) $\sim$
 1.44 has been estimated for the nebula \citep{tsamis03}.
 The NUV filter, N279N has a passband of \SI{90}{\angstrom} and includes \ion{Mg}{ii} at \SI{2800}{\angstrom} \citep{feibelman01,beintema99}, and [\ion{Mg}{v}] at
 \SI{2786}{\angstrom} \citep{barral82} emission lines. 
 
                 In  Fig.~\ref{fig:iuespectrum} one of the offset (red) spectra has been shown
 scaled to match the [\ion{N}{iv}] at \SI{1486}{\angstrom} emission line in the nebular centre
 spectrum (blue) to emphasize how the emission lines vary away from the centre (star). 
 The strength of the emission lines seem to decrease away from the centre. Moreover,
 the high excitation emission lines like \ion{C}{iv} at \SI{1550}{\angstrom} weaken much more than
 the moderate excitation lines. A similar trend is seen with the other
 offset spectrum (black) as well.
  The stellar \ion{C}{iv} at \SI{1549}{\angstrom} feature is weakest (as
would be expected), while the nebular \ion{N}{iii}] at \SI{1750}{\angstrom} emission
is the strongest in the offset spectra. \ion{He}{ii} is less
affected, suggesting that it may be a mixture of stellar and
nebular emission.

\begin{figure*}
\centering
\includegraphics[width=\textwidth]{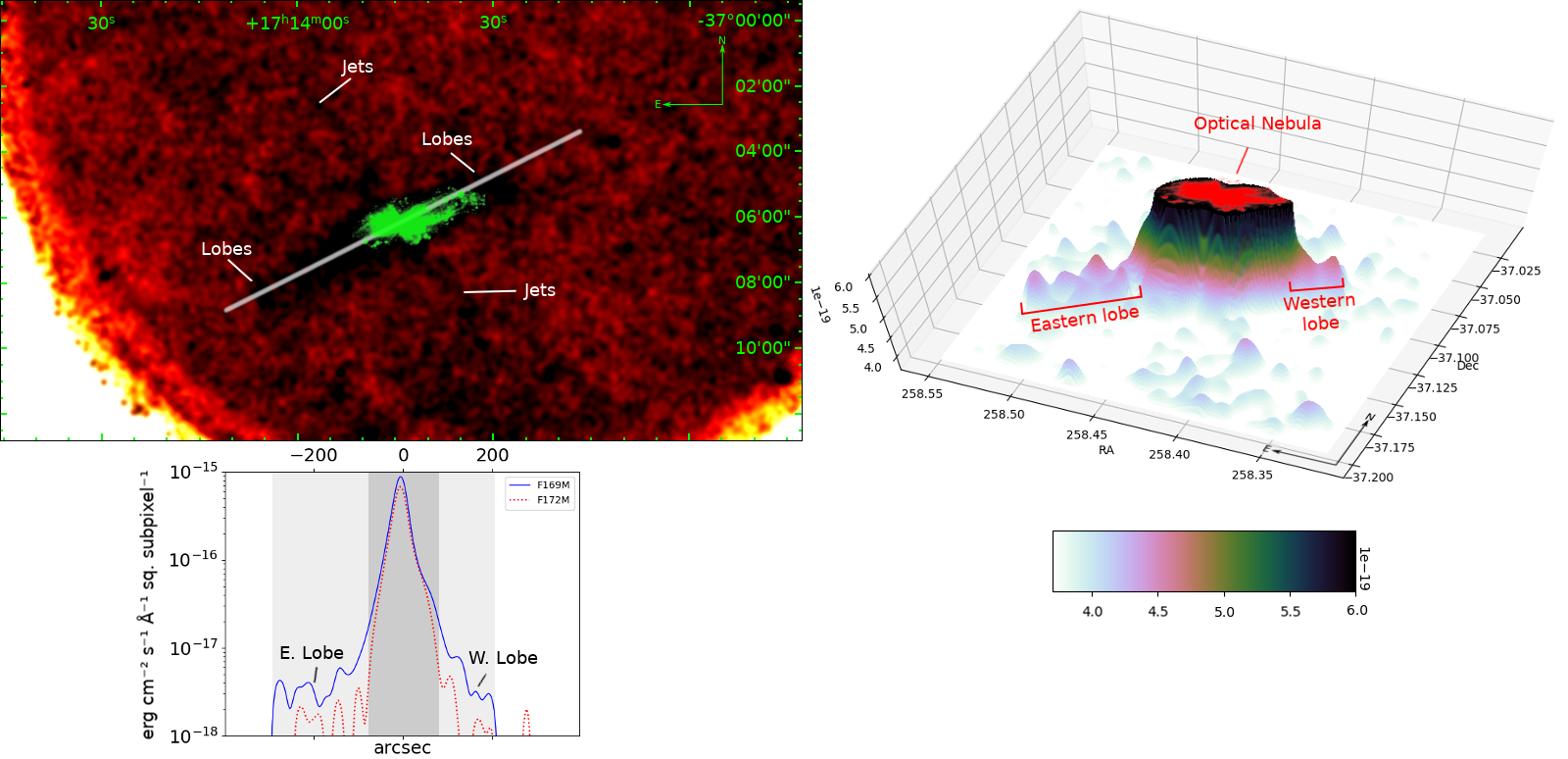}
\caption{ Top left: enhanced UVIT image (smoothed and with a higher contrast to show larger and fainter features) of \object{NGC~6302} in F169M filter showing 
the fainter
 extensions. The  FUV lobes are 
  inclined at an angle to the east-west direction at PA of 113$^\circ$ .
  The `jets' are almost perpendicular to the FUV lobes.
 An optical image has been superposed (in green) on the figure for comparison of the size of FUV lobes and jets with the optical nebula. Note that
 the optical lobes are extended in east-west direction and the torus  in
 north-south. (bottom): we show the profiles of crosscuts (in flux density 
 units - \si{\fluxdensityunitperpixel} across the white line shown in top left panel) in the filters F169M and F172M.  Note the extent of the lobes in the
 F169M image (blue) relative to that of the F172M image (red).
 Although the core of the profiles in both filters  
 are of same extent and are similar to that of optical nebular images, only F169M image has lobes that extend on either 
 side to about 5~arcmin. (top right): A star‑subtracted 3d plot of the F169 image. An optical image(in red) has been superposed for comparison of extent. The eastern and shorter western lobes can be clearly seen in this plot.
 }
\label{fig:lobes}
\end{figure*}
\section{Results}
\label{sec:res}
\subsection{Morphological Description }
\label{ssec:morphology}

      In Fig.~\ref{fig:imagecomp}  we present a comparison of images of \object{NGC~6302} in all four
 UVIT filters. UVIT
 offers a better spatial resolution than Galex.
 The image in N279N (Fig.~\ref{fig:imagecomp} - top left) tracks the
 \ion{Mg}{ii} \SI{2800}{\angstrom} line emission and is somewhat similar to the
 [\ion{N}{ii}] (and [\ion{S}{ii}]) images shown in \citet{hua98} and 
\citet{meaburn05}, but is  more extended to the west.

The  UVIT N219M image (Fig.~\ref{fig:imagecomp}; top right) is centred on the $\lambda$ \num{2200} ISM absorption band. No strong emission lines are present in the 
 \object{NGC~6302} IUE spectrum in this passband. The image is similar to the Galex NUV but the western lobe seems to be
 much fainter suggesting extinction in that region.

    In Fig.~\ref{fig:imagecomp}, bottom panels,  the UVIT FUV images are shown.
  The F172M-filter image (bottom  right) includes \ion{N}{iii}] \SI{1760}{\angstrom} emission and
 looks very similar to the [\ion{O}{iii}] (and H$\alpha$) images from 
 \citet{hua98}.
 \citet{hua98} quote the size of
  the nebula in various emission lines including 
 [\ion{O}{iii}] \SI{5007}{\angstrom} which  has dimensions of $\SI{60}{\arcsecond}\times \SI{185}{\arcsecond}$. 
 The image in F169M filter, which includes high excitation lines (Fig.~\ref{fig:iuespectrum}),
 is expected to be of the same or smaller in size  than the [\ion{O}{iii}] image. The inner part of the nebula shown in Fig.~\ref{fig:imagecomp} (bottom left) of  F169M  image has
 similar size to the [\ion{O}{iii}] image and and displays the high excitation
  line region.

  The enhanced UVIT F169M filter image, however, (Fig.~\ref{fig:lobes}, top left) shows  a different size and
  appearance. Surprisingly it shows major extensions that are not present in the optical nebula. The western lobe
  extends  to the north-west like a band with constant width (further discussed in Sect.~\ref{ssec:lobesandjets}).

 Both NUV N279N and FUV F169M images show traces of bow-like filamentary structures about 4.5~arcmin west of the nebular centre. However the intensities of these filaments are marginal and cannot be  reliably distinguished from the background without further observations.

\subsection{FUV lobes and jets}
\label{ssec:lobesandjets}

        The most surprising discovery of our UV observations is the large, but faint, extended
 lobe  emission (much larger than the optical lobes) in the deeper representation
  of F169M-filter image (Fig.~\ref{fig:lobes}). In addition there are two thin, collimated features that we shall refer to as ``jets'' extending almost 
  perpendicularly  to the FUV  extended lobes 
  through the centre of the nebula on both sides. The lobes and jets are  present
  only in the F169M filter
 images but not  in the FUV F172M image nor in the UVIT and Galex NUV filters. A slightly smoothed negative F169M image is shown in
 Fig.~\ref{fig:lobes} (top left), where the  FUV lobes and jets are compared with the optical [\ion{N}{ii}]
  image from \citet{meaburn05}.

  A crosscut (\si{\fluxdensityunitperpixel} vs size) across the 
  images of F169M and F172M (Fig.~\ref{fig:lobes}, bottom)
 shows the extended emission and the presence of lobes in the F169M image compared to the
 F172M image, although the nebular core emission is similar. The lobes are at a 2~$\sigma$ level above the background. The in-orbit flux 
 calibration from \citet{tandon17b} has been used in constructing the flux 
 profiles.   It appears that  FUV lobes  and jets only radiate
 in the F169M band and at shorter wavelengths.

             How real are the FUV lobes and jets?
  There are no known instrumental artefacts that could create such a pattern. The
  diffraction pattern expected from the secondary mirror support system is 
  quite different (it has eight uneven ribs) and is not seen even in images of bright stars.

  In addition to the the F169M image 
 we have a short exposure (although we specified
 a long exposure of 1500s, the satellite operations for some unknown reason 
 only provided 540s 
 exposure out of which only 400s was useful) of the nebula through the  F154W filter (BaF${\rm 2}$).
 The transmission 
 of the F154W filter is similar that of F169M
 but with a broader bandwidth of \SI{380}{\angstrom} and $\lambda_{\rm eff}$ of \SI{1541}{\angstrom} 
 extending shortward to \SI{1340}{\angstrom} \citep{tandon17b}. The F154W image shows
  the presence of FUV lobes and a trace of the northern jet at the
  same locations as in the F169M image.
  Thus the FUV lobes and  jets  exactly located at the centre of
  the nebula (and the star) appears to be real and significant. 
 
         The deep [\ion{N}{ii}] \SI{6583}{\angstrom} image of \citet{hua98} displays
 a  bright jet or a filament in north-north east edge (Fig. \ref{fig:deepn2}) that coincides with the base of the northern 
 FUV jet. It may be no coincidence that they are located almost
  on top of each other.

               The FUV lobes are  oriented at an angle of about \ang{23} to the optical lobes.
  The optical lobes are essentially oriented in the east-west direction with
  the dark lane torus in the north-south (within $\pm$\ang{2}). The FUV lobes, on
  average, are at position angle of 
  \ang{113} $\pm$ \ang{1}.
  The mean jet orientation has a position angle of \ang{35}  (hence about
  \ang{12} from the axis perpendicular FUV lobes - not exactly 
  orthogonal).
   The total extent of the FUV lobes is about 10.25$\pm$0.1~arcmin (about
  5.12~arcmin from the central bright region). The width of the
   eastern part is about 1.33~arcmin. The jets extend to a total length
   of 9.15~arcmin with 4.55~arcmin on either side of the FUV lobes with a width of
   about 18$\pm$2~arcsec. The FUV lobe  thickness seem to be more or less 
  constant across the length (until the edge of NW side - the  western 
  part may be  slightly more tapered and less thick),  suggesting that it is
  a result of either a collimated flow or a disc of constant thickness.

             At a  distance of 1.17 kpc, the FUV lobes  have a radius of 
 \SI{5.4e18}{cm} (\SI{1.74}{\parsec})  with a thickness (width) of \SI{1.4e18}{cm} (\SI{0.45}{\parsec}). The jets extends to a physical distance of \SI{4.8e18}{cm} (\SI{1.55}{\parsec})
  with a thickness of \SI{3.1e17}{cm} (\SI{0.1}{\parsec}) on each side.
 
    These dimensions are quite different from those of the optical lobes.
  \citet{meaburn05} find the NW lobe  extending to about 3.9~arcmin
  (\SI{1.33}{\parsec}) from the centre and the eastern lobe 
  seems to 
  extend to about 1.5~arcmin (\SI{0.51}{\parsec}) from the centre \citep{szyszka09}.
  The torus, consisting of two rings, extends to a distance of \SI{9e16}{cm} (inner ring) to \SIrange{0.95e17}{1.2e17}{cm} (outer ring) from the centre \citep{santandergarcia17}.
 
                \citet{huggins07} discusses the origins of jets and tori in young PN
 and pre-PNs. He lists the dimensions of jets and tori around some of the
  typical systems. The FUV disk and jets of \object{NGC~6302} are much larger than
  those listed by \citet[may be instead comparable to the \object{KjPn~8} system]{huggins07}.

\iftrue
\begin{table*}
\centering
\caption{ Observed flux of NGC~6302 in various filter bands  }
\begin{tabular}{lrrr}
	\hline \hline
	Filter/line                           &             Core   &           Lobes   &            Jets   \\ 
	                                      &  (\si{\fluxunit})   & (\si{\fluxunit})   & (\si{\fluxunit})   \\
    \hline
	F169M                                 &    \tablenum{4.1e-11}   &   \tablenum{1.2e-11}   &   \tablenum{1.7e-12}   \\
	F172M                                 &    \tablenum{1.7e-11}   &                   &                   \\
	F219M                                 &    \tablenum{3.0e-12}   &                   &                   \\
	N279N                                 &    \tablenum{4.9e-12}   &                   &                   \\
	F154M                                 &    \tablenum{5.8e-11}   &              --   &              --   \\
	F(H$\beta$)$^{a}$                     &   \tablenum{2.07e-11}   &                   &                   \\
	F(H$\alpha$)$^{a}$                    &   \tablenum{1.77e-10}   &                   &                   \\
	F([N\,{\sc ii}]$\lambda$6583)$^{a}$   &   \tablenum{4.53e-10}   &                   &                   \\ 
    \hline
    \multicolumn{4}{l}{$^{a}$ From \citet{hua98}}                                                             \\
\end{tabular}
\label{tab:flux}
\end{table*}   
\fi

\begin{figure*}
\centering
\includegraphics[width=\textwidth]{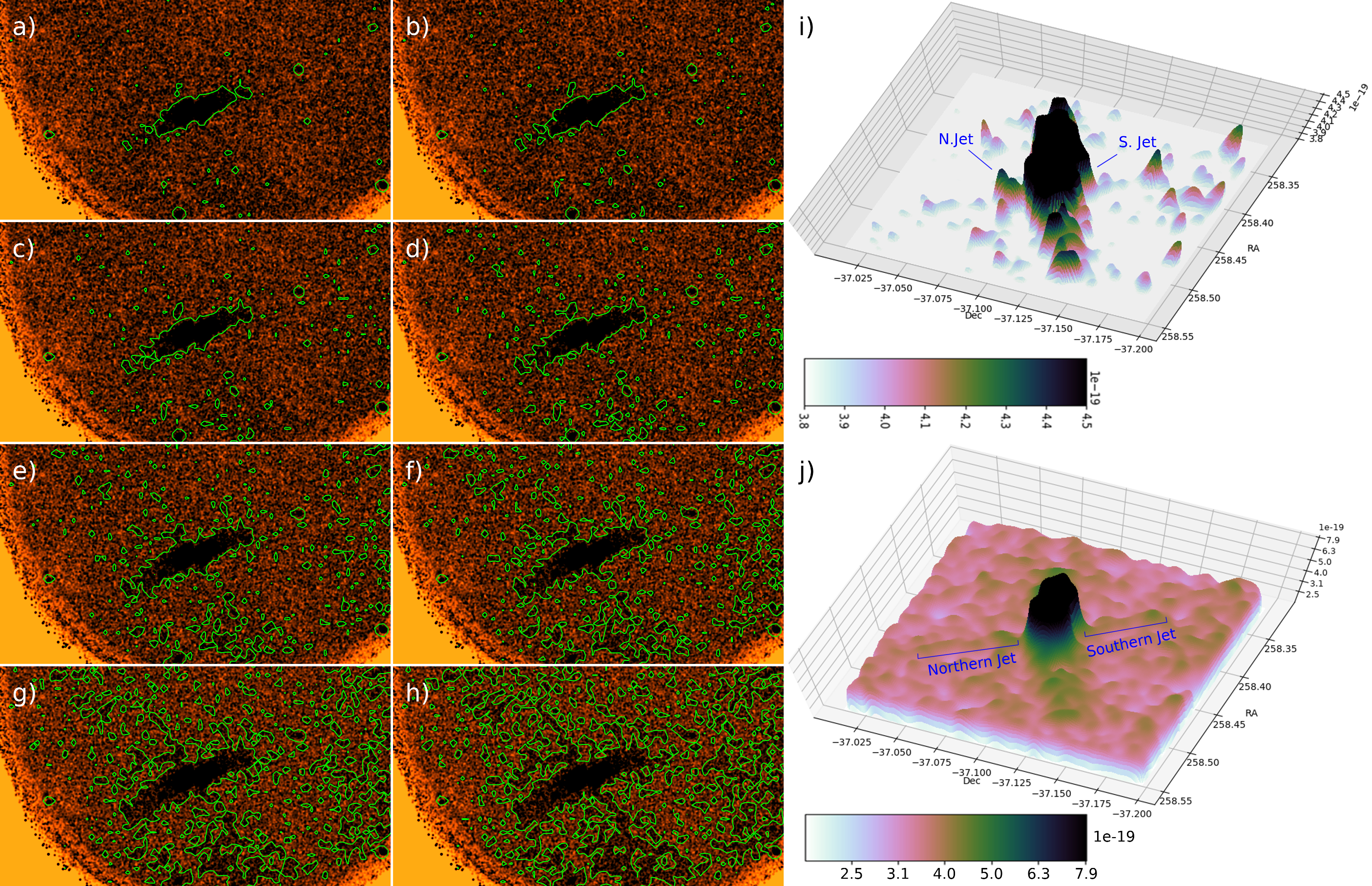}
\caption{ (a-h): contour plots of the F169M image at various flux levels. The contours in these figures correspond to a flux above background of
a) 2.2~$\sigma$
b) 1.9~$\sigma$
c) 1.7~$\sigma$
d) 1.5~$\sigma$
e) 1.2~$\sigma$
f) 1.0~$\sigma$
g) 0.7~$\sigma$
h) 0.5~$\sigma$.
In c) we note the beginnings of the jet both to the North as well as the South of the nebula. As we move towards the edge of the jets the flux level falls off with the full extent being visible in (g-h). From (e-f) we note that the jet is not homogeneous in flux. (i): star-subtracted 3d plot in the flux range mentioned above. The small-extensions to the North and South the ones seen in c) and in Fig. \ref{fig:deepn2}. (j): similar to the plot above but with a larger range in flux. The entire jet can be seen as a ridge similar to what we see in (g).
 }
\label{fig:con3d}
\end{figure*}

\begin{figure}
\centering
\includegraphics[width=\columnwidth]{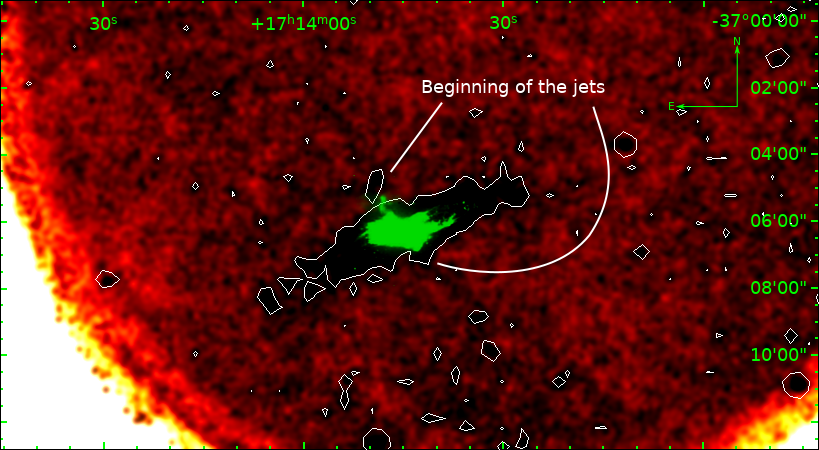}
\caption{The enhanced F169M image with contours at \SI{4.2e-19}{\fluxdensityunitperarcsec} (1.7~$\sigma$; Fig. \ref{fig:con3d}; (c)) superposed (in green) with a deep [\ion{N}{ii}] image from \citet[Fig. 18 (b)]{hua98}. We note the similarity in shape of the extensions to the North and South of the nebula in the both F169M and [\ion{N}{ii}] image, though in the F169M image the features are of a larger size.
 }
\label{fig:deepn2}
\end{figure}

The contour plots (Fig. \ref{fig:con3d}; (a-h)) shows gradual change in the size of the jets and lobes with flux. The spatial extent of the structures is presented in Table~\ref{tab:extent}, where we list the size as a function of flux level above the background noise. At higher flux levels (2~$\sigma$; Fig. \ref{fig:con3d} (c)), the jet size is about 3.6~arcmin. As we start looking at lower flux levels the jet size increases. As we approach the level of the background (0.7~$\sigma$; Fig. \ref{fig:con3d} (c)) we can see the full extent of about 9.0~arcmin.

\begin{table*}
\centering
\caption{Extent of the lobes and jets above various flux levels (background subtracted). The corresponding sigma level are also mentioned.}
\label{tab:extent}
\begin{tabular}{lrrrr}
\hline \hline
Flux density                    & E. lobe & W. lobe & N. jet & S. jet \\
\si{\fluxdensityunitperarcsec}  &  arcmin &  arcmin & arcmin & arcmin \\
\hline
\num{5.4e-19} (2.2~$\sigma$)    &     3.5 &     3.4 &    1.9 &    1.5 \\

\num{4.2e-19} (1.7~$\sigma$)   &     4.2 &     3.7 &    2.4 &    2.1 \\

\num{3.0e-19} (1.2~$\sigma$)   &     4.8 &     3.8 &    3.7 &    2.8 \\

\num{1.8e-19} (0.7~$\sigma$)   &     5.1 &     3.9 &    5.3 &    3.6 \\

\hline
\end{tabular}
\end{table*}

 Using the in-orbit flux calibration given in \citet{tandon17b}, we
 estimated the total flux in FUV lobes, jets and the core of the nebula in
 all UV filters (with help of contoured images) as shown in Table \ref{tab:flux}.  The exposure of
 in  F154W filter is too weak to properly estimate the flux in the FUV lobes and jets.
  These fluxes can be compared
 with the total nebular fluxes in the optical wavelengths as estimated by
 \citet{hua98}. The UV flux estimates have an
 uncertainty of 10 to 15 percent.

\section{Discussion} 
\label{sec:disc}

\subsection{What is the source of FUV emission in the large-scale lobes and jets}

            The source of FUV emission in the outer parts of the FUV lobes and
  the jets cannot be attributed to high excitation emission  lines
  (shown in Fig.~\ref{fig:iuespectrum}), which are the dominating source close to the central star
   and the region occupied by the optical lobes as exemplified by the [\ion{O}{iii}]
   images \citep{hua98}.
  Scattering of the stellar
  spectrum by nebular dust (Mie scattering)  is also unlikely, because scattering
  would be over a wide wavelength range and
  images in the F172W filter, that is centred about \SI{100}{\angstrom}  longward of F169M
 (as well as  NUV filters) do not show the FUV lobes nor the jets as would be instead expected.
 
             The other possibility for the FUV F169M filter emission is
  through the Lyman bands of the
 H$_{\rm 2}$ molecule. H$_{\rm 2}$ from circum-nebular
  regions
 has  been detected around several PNs \citep{herald11},
   reflection nebulae \object{IC~63} \citep{witt89}, and in FUV rings around
    the carbon stars \object{U~Hya}
 \citep{sanchez15}, IRC+10216 \citep{sahai10} and CIT 6 \citep{sahai14}. UV fluorescence spectra of H$_{\rm 2}$, as modelled by
  \citet{france05} of \object{IC~63} show strong emission blueward of \SI{1608}{\angstrom} peak,
 and no emission longward of \SI{1650}{\angstrom}. Werner and Lyman bands of H$_{\rm 2}$
  extend up to about \SI{900}{\angstrom}. Since the FUV lobes and jets in \object{NGC~6302}  are
  only present in images using the
        F169M filter, but not in those using the
   F172W filter (or in NUV filters),
           this suggests that H$_{\rm 2}$ emission might be the source. Although weak, the
  presence of lobes and jets in the  F154M images provides additional support for this suggestion.
  A similar conclusion was arrived at for the FUV halo emission in \object{NGC~40} \citep{rao18}. Although the central star of \object{NGC~6302} is very hot and luminous,
  it is still  surrounded by 8 mag of visual extinction (as viewed from Earth),
  due to the dusty torus. Whether sufficient stellar UV flux reaches the ends of the FUV lobes
  and  the jets so as to provide H$_{\rm 2}$ fluorescent emission, is
  not obvious, but is possible. \citet{wright11} estimate that a large 
  fraction of the central star's radiation would be escaping through the polar axis
  (orthogonal to the torus).
   UV extinction through the optical lobes might still
  not be large enough to block the star light.

          Collisional excitation of H$_{\rm 2}$ by hot electrons has been 
 suggested as the source for FUV emission rings seen around carbon
 stars \object{IRC~10216} and \object{CIT~6} \citep{sahai10,sahai14}.
 The central stars are too cool. Bow shocks generated
  from the interactions of cool molecular stellar winds with the ISM  provide
  the hot electrons needed to generate FUV emission.   

     Collisional excitation of H$_{\rm 2}$ by hot electrons is not  very
  likely in the case of \object{NGC~6302} since no real evidence for strong shocks
  is seen. Moreover FUV emission in \object{NGC~6302} is  morphologically
  different from the FUV rings around carbon stars. The FUV emission comes
  from the filled-in lobes and jets in \object{NGC~6302}. Hot electrons generated in
  the shocked regions (stellar wind interaction with the ISM?) have to fill the
  entire volume of FUV lobes and jets, not only the bow shock regions. Such a
  major source for the production of electrons is not obvious. In such a case
  it is surprising that no collision line emission (e.g., say from [\ion{O}{i}])
  is seen from the FUV lobes in the optical spectral range. In case of PNs the central stars
  are hot enough to provide the FUV radiation for the H$_{\rm 2}$ fluorescence
  to be operative.

  H$_{\rm 2}$ rotational lines in the infrared region from the lobes have been
  reported by \citet{beintema99} in their ISO spectrum. \citet{latter95} imaged the nebula in H$_{\rm 2}$ at
  \SI{2.121}{\micro\metre}, which showed a very similar appearance as the Brackett $\gamma$ line
 image and no extension similar to the FUV lobes or jets.

      Assuming that the FUV lobes and jets emission is due to H$_{\rm 2}$
 fluorescence, the total emitted flux from the nebula through the lobes and the jets  is estimated as
  follows. Earlier (in Sect.~\ref{ssec:lobesandjets}) we estimated the observed flux in the FUV lobes and jets through the F169M filter.
  The interstellar extinction towards NGC~6302 has been estimated to be \mbox{E(B-V) =  0.99~mag}. Correcting the
 fluxes for interstellar extinction using Seaton's reddening curve \cite{seaton79} the reddening-free 
  fluxes are \SI{1.72e-8}{\fluxunit} and 
 \SI{2.4e-9}{\fluxunit} for lobes and jets, respectively. The total
  fluxes emitted by \object{NGC~6302} located at a distance of \SI{1.17}{\kilo\parsec} are
 \SI{2.85e36}{erg~s^{-1}} and \SI{4.0e35}{erg~s^{-1}} in the F169M filter for lobes and jets respectively.
 It has been generally estimated, in case of H$_{\rm 2}$ fluorescence, that 30 percent of
  the flux is emitted in Lyman bands in the wavelength range of
   \SIrange[range-phrase={ to }]{1400}{1700}{\angstrom} \citep{martin90},  
 which is appropriate for the F169M filter. The rest of the emission is expected to be
  in the Werner bands between \SI{940}{\angstrom} and \SI{1300}{\angstrom}. Thus   the total
  H$_{\rm 2}$ fluorescent
  emission  by \object{NGC~6302} in FUV lobes and jets is estimated to be 
  \SI{1.0e37}{erg~s^{-1}} and \SI{1.3e36}{erg~s^{-1}}, respectively.
  The uncertainty in these estimates is about 30 percent.\footnote{Gomez (2013) causally suggests neutral C excitation through inelastic collisions
  of  $^{3}P$- $^{1}D$ transitions might provide FUV emission - but not sure
 that this would work here (might contribute to a minor extent?)}

  \citet{hsia14} present [\ion{N}{ii}] images of young multipolar nebulae
  \object{Hen~2-86} and \object{Hen~2-320}  (their class II type) that show two and four pairs of 
  collimated lobes respectively with the same bipolar axis. 
  The ends of each pair of these  lobes 
  are  well  marked by brighter nebular arcs (rings) perpendicular to the lobe axis.
  Sometimes the inner lobes have larger width relative to the outer ones.
   Although the origins of these multilayer bipolar lobes is unclear,  one
  possibility suggested by \citet{hsia14} is consecutive ejections collimated 
 by similar 
  mechanism.

\subsection{How are FUV lobes related to the optical nebula?}

\citet{meaburn08} determined an age of NW {\it optical} lobe of \SI{2200}{\year} from proper motion studies. By modelling the 
 spatio-kinematic structure observed in the HST images using the SHAPE program, \citet{santandergarcia17} estimated
  the kinematic age of the wide lobes to be between \SI{3600}{\year} and \SI{4700}{\year}. On the
  other hand, from  proper motion studies of the HST images of the eastern 
 lobe, \citet{szyszka11}  determined an age of \SI{2200}{\year}, in line with 
  the conclusion of \citet{meaburn08}, from which one can conclude that
 the Hubble-type flow seen in both eastern and NW lobes might be the result of
 the same explosive event triggered $\sim$\SI{2200}{\year} ago.
               
 The massive molecular ring (not seen in the optical images), that constitutes the torus, was
 formed between 3000 and 5000 years ago, according to the modelling studies of 
  \citet{santandergarcia17}, after which the perpendicular optical 
 lobes were  ejected. According to their scenario, two major ejection events
  have happened to the AGB star that produced \object{NGC~6302}: the formation of the
  torus and  the formation of the lobes (the Hubble flow). The 
  photoionisation models of \citet{wright11} suggest that central star
  left the AGB stage some \SI{2100}{\year} ago, in agreement with the time of the 
 optical lobe ejection event.   

While it is not easy to find a perfect scenario to fit our knowledge of the optical nebula, it is likely that a binary interaction ejected a large circumbinary disk (the molecular torus) during a phase of unstable Roche lobe overflow. That phase then proceeded to a common envelope event, possibly even one resulting in a merger and it is that event which produced a second ejection interacting with the disk. The overall shape would then have been ploughed by the fast wind, ionised, and resulted in the optical nebula. Since quantitative theoretical knowledge is scarce \citep[see for example][]{nordhaus06}, this scenario is plausible. The story, however, becomes far more complex, even for a qualitative scenario, when we add the information provided by the large FUV structures.

                Since we do not have any kinematic  information of the FUV 
  lobes and jets we cannot estimate their age.
  However, solely  based on their larger size they are likely older than the optical nebula.  
 
       The typical velocity
  of jets in bipolar systems is about \SIrange{100}{200}{km.s^{-1}} and 
  tori are known to expand at about \SI{10}{km.s^{-1}} \citep{huggins07}. If we assume that a velocity 
  of \SI{160}{km.s^{-1}} characterises the FUV lobes as well as the jets,
  their  age would be \SI{10600}{\year} and \SI{9500}{\year}, respectively. 
   This estimate, although uncertain, is  much larger
  than the ages of the optical lobes and of the 
  torus and would indicate that the FUV lobes and jets  
  represent an earlier event.

The orientation of the FUV lobes and jets with respect 
  to the axis of optical nebula may present further clues (Fig.~\ref{fig:lobes}).  
   Next we note that the axis of the FUV lobes is not the same as the long axis of the optical nebula (it is tilted $\sim$23$^\circ\pm$1$^\circ$ anticlockwise with respect to the long axis of the optical nebula), although one could argue that the NW FUV lobe seems aligned with the longest protrusion of the optical western lobe.
 
Another feature of the the FUV lobes (and particularly of the eastern 
lobe) is that they have the same width throughout their length, suggesting a certain type of collimation. 
  On the other hand, the optical lobes have a wide opening angle and have a Hubble-type characteristic with a radial expansion from the central star (shown by the proper motion vectors of Fig.~9 in \citet{szyszka11}). It is
  highly unlikely that the FUV lobes are a similar Hubble-type flow.  

These characteristics are not immediately explainable, though taken in groups they can be found in other nebulae. For example several quadrupolar and multipolar nebulae  have been suggested as examples where
 a change of orientation in the ejection \citep{manchado96}. These
  lobes are thought to be ejections at different times in the course of a
  precession cycle of the rotation axis of the central star in a binary
  system. \citet{manchado96} estimated, in \object{PN~M2-46}, that the rotation axis has
 changed by \ang{53} between two ejections of the lobes (see their Fig.~5)
In the case of NGC~6302, the axis seems to have changed by \ang{67}.  Similarly, the ejection type is clearly different (radial Hubble-type flow vs. cylindrical and likely not Hubble-type). 
 
 At what stage in the evolution of an AGB  binary 
these ejections take place is not clear. If we assume that the optical nebula and the molecular torus were ejected one after the other at the of the AGB life of the progenitor, the FUV structures would have been ejected earlier, but why? Could the system have been a triple where the older ejection (FUV structures) was at the hand of an inner companion (a merger that did not terminate the AGB life of the star) and the later ejection at the hand of a companion farther out, reached by the continually expanding AGB star? And if so, how was the older ejection accomplished, how can two sets of perpendicular lobes and jets be generated? 

Another way to produce two distinct interaction events, and possibly two
nebulae, at a distance of a few thousand years could be related with a
post-AGB final helium shell flash \citep{demarco08}. This would be also in
line with the hydrogen-deficient nature of the central star \citep{wright11}: 
a common envelope event took place that terminated the life
of the AGB star (first ejection producing the FUV nebula) and a close
binary star surrounded by a nebula; then a second outburst, a last
helium shell flash, would have likely lead to the expansion of the
primary, leading immediately to a  merger with the close companion and
possibly the second (optical) nebula. This is a plausible scenario since
post-CE binaries exist in the core of PN \citep{hillwig16} {\it and}
we know that about 20 percent of all post-AGB star suffer a last helium
shell flash \citep{werner06}. This scenario is qualitative and
does not necessarily explain all the data - the orientation of the
ejecta remains a mystery as does the reason why the outer nebula is so
faint and only seen in FUV. However, its broad characteristics are in
line with a double outburst, something that is generally difficult to
obtain.

  Although the origin of collimated outflows and jets is quite uncertain,
    most often they are linked to interactions within binary systems 
   involving magnetic fields and accretion processes \citep{soker98,soker06,tocknell14,soker94,reyesruiz99,nordhaus06,blackman14}. 
   In some cases these ejection could be quite ``explosive'', generating  Hubble-type
   ejecta.
  \citet{soker12} invoke similarities in the  properties of optical lobes and
  torus of \object{NGC~6302} with `intermediate-luminosity optical transients' 
  (ILOT) which are explosive events  
  with energies intermediate between novae and supernovae \citep{kasliwal11}.
  
The FUV lobes and jets of \object{NGC~6302} seem to share some characteristics with
   young multipolar collimated lobe systems  like those in \object{Hen~2-320}, \object{OH~231.8+4.2}, or \object{IC~4406}
  but are \numrange{5}{10} times larger in size. Clearly the diversity of initial conditions offered by binary and triple systems makes for a maze of possible structures that, particularly in light of an incomplete theoretical understanding of mass transfer, makes for a very confusing landscape.

\section{Concluding remarks}
\label{sec:concl}

    UVIT images of the 
   young butterfly nebula \object{NGC~6302},  revealed dramatic large FUV lobes and orthogonally placed jets that are 
   unseen optically. FUV lobes are much larger than the optical lobes.
   These new structures add to the complexity of structures already seen in the
   optical.
   Do FUV and optical  structures need
  a binary central star?
  The source of FUV emission shortward of the 
   F169M filter bandpass  is not known but is most likely
  fluorescent emission of the H$_{\rm 2}$ molecule. Substantial mass might be
   hidden in the unseen H$_{\rm 2}$. UV absorption line (and extinction)
  studies of background stars through these FUV lobes might help in 
  estimating their total mass. Deep far IR images might show cold dust 
  associated with these FUV lobes.

           Such FUV emission structures may not be uncommon features of young
  PNs . UVIT images of NGC 40 already showed a FUV halo \citep{rao18}.
   Another bipolar (multipolar) nebula, \object{NGC~2440} also shows FUV extensions
    and a jet
   beyond the optical (and near UV) nebula (Kameswara Rao {et~al.} in 
  preparation). UVIT on
  ASTROSAT with its angular resolution ($\sim$ \SI{1.4}{\arcsec})  comparable to ground
  based imagery and high sensitivity is well suited to map such FUV structures
   so far hidden.

\begin{acknowledgements}
  UVIT and ASTROSAT observatory development  took about two decades before launch. Several people from several agencies were involved in this effort. We would like to thank them
 all collectively. NKR and KS would like thank Department of science and
  technology for their support through grant
  SERB/F/2143/2016-17 `Aspects in Stellar and Galactic Evolution'. AR acknowledges
 Raja Ramanna Fellowship of Department of Atomic energy for the support.
 NKR would like to thank Dr.George Koshy for his help in estimating the total
 flux in the lobes.  

 Some of the data presented in this paper were obtained from the Mikulski 
 Archive for Space Telescopes (MAST). STScI is operated by the Association of
 Universities for Research in Astronomy, Inc., under NASA contract NAS5-26555.
 Support for MAST for non-HST data is provided by the NASA Office of Space
 Science via grant NNX09AF08G and by other grants and contracts.

\end{acknowledgements}


\end{document}